\documentclass{article}
\usepackage{ijcai26}

\usepackage{times}
\usepackage{soul}
\usepackage{url}
\usepackage[hidelinks]{hyperref}
\usepackage[utf8]{inputenc}
\usepackage[small]{caption}

\usepackage{amsmath}
\usepackage{amsfonts}
\usepackage{amsthm}

\usepackage{algorithm}
\usepackage{algorithmic}
\usepackage[switch]{lineno}

\usepackage{adjustbox} 

\usepackage{array} 
\usepackage{multirow}
\usepackage{graphicx}
\usepackage{booktabs}
\newcommand{\matr}[1]{\mathbf{#1}}

\newcommand{\citet}{\cite}

\newcommand{\Xm}{\matr{X}}
\newcommand{\sasrec}{SASRec~}

\usepackage[dvipsnames]{xcolor}
\newif\ifdraft
\drafttrue   

\ifdraft
  \newcommand{\ef}[1]{\textcolor{blue}{[EF: #1]}} 
  \newcommand{\tn}[1]{\textcolor{orange}{[TN: #1]}} 
\else
  \newcommand{\ef}[1]{}
  \newcommand{\tn}[1]{}
\fi

\title{Position-Aware Sequential Attention for Accurate Next Item Recommendations}

\author{
Timur Nabiev$^1$
\and
Evgeny Frolov$^{2}$
\affiliations
$^1$Skolkovo Institute of Science and Technology\\
$^2$AIRI\\
\emails
Timur.Nabiev@skoltech.ru,
frolov@airi.net
}

\begin{document}

\maketitle

\begin{abstract}
Sequential self-attention models usually rely on additive positional embeddings, which inject positional information into item representations at the input. In the absence of positional signals, the attention block is permutation-equivariant over sequence positions and thus has no intrinsic notion of temporal order beyond causal masking. We argue that additive positional embeddings make the attention mechanism only superficially sensitive to sequence order: positional information is entangled with item embedding semantics, propagates weakly in deep architectures, and limits the ability to capture rich sequential patterns. To address these limitations, we introduce a kernelized self-attention mechanism, where a learnable positional kernel operates purely in the position space, disentangled from semantic similarity, and directly modulates attention weights. When applied per attention block, this kernel enables adaptive multi-scale sequential modeling. Experiments on standard next-item prediction benchmarks show that our positional kernel attention consistently improves over strong competing baselines.
\end{abstract}

\section{Introduction}\label{sec:intro}
Sequential recommendation aims to predict a user’s next interaction based on their recent behavior, typically represented as an ordered sequence of items. Self-attention models such as \sasrec have become a standard choice for this task, as they can flexibly aggregate information from all past positions in a sequence. To make attention sensitive to order, these models usually rely on additive positional embeddings that are added to item representations at the input layer.

Despite their empirical success, additive positional embeddings present an ad-hoc way to indirectly inject positional information into the attention mechanism. In the absence of any positional signal, the self-attention block is permutation-equivariant over sequence positions, meaning that reordering the items only permutes the outputs and the block has no intrinsic notion of temporal order beyond causal masking. Additive positional embeddings break this symmetry by mixing position vectors with item representations, but they do so only at the input embedding layer, leaving the attention operator itself structurally unchanged.

This design has two important consequences. First, positional information is entangled with item semantics, so attention weights cannot distinguish whether a correlation arises from content similarity or from positional structure. Second, because positional signals are injected only once and then propagated through multiple layers, the sensitivity of deep models to sequence order can become weak and difficult to control, limiting their ability to capture richer multi-scale sequential patterns.

In this work, we propose an alternative approach to positional modeling in self-attention. Instead of encoding positions additively at the embedding level, positional information is introduced multiplicatively inside the attention operator through a learnable position-aware kernel. This kernel operates purely in the position space and is explicitly disentangled from content similarity. It directly modulates attention weights while preserving causality by restricting interactions to past positions.

More precisely, the attention mechanism is augmented with a positional kernel matrix that is factorized into two triangular components, which respectively control how positional structure affects attention scores and how it influences the aggregation of value vectors. The kernel can be parameterized per attention block, allowing different layers to specialize to different temporal dynamics and potentially making the overall model more expressive in capturing various short- and long-range dependencies. 

Compared to standard self-attention with additive positional embeddings, the resulting position-aware kernel attention offers a more flexible parameterization of temporal effects by modeling them directly through an explicit positional kernel inside the attention operator, while remaining efficient and easy to integrate into existing architectures. The main contributions of this work are threefold.
\begin{itemize}

    \item We provide a focused analysis of positional modeling in sequential self-attention and argue that the standard additive positional encoding scheme is a weak and indirect way to capture sequence order in deep architectures.
    \item Building on a formal analysis of the well-known permutation-equivariant nature of self-attention without positional signals, we identify several mechanisms that contribute to this weakness -- early additive fusion with item embeddings, entanglement of positional and semantic information, and lack of explicit control over positional effects. We use them to motivate an alternative, kernel-based positional scheme.
    \item Based on these findings, we introduce a learnable position-aware kernel that resolves permutation equivariance directly inside the attention operator, operates purely in the position space, and is disentangled from semantic similarity of items.

\end{itemize}

The source code to reproduce our work is available online\footnote{Anonymous repository for review \url{https://github.com/anonymcode/IJCAI2026}.}.

\section{Background and problem formulation}\label{sec:preliminaries} 

In this section, we briefly introduce main self-attention components and analyze them as a composition of shared linear operators acting over a matrix representation of sequences. It helps us to reveal how positional equivariance arises and why additive positional encodings are structurally weak. It further motivates the structural design of our position-aware kernel.

\subsection{Sequential self-attention}\label{sec:attention}

Consider a catalog of $N$ items with each item represented by a one-hot vector in $\mathbb{B}^N$. A user's item sequence can be represented as matrix \mbox{$\Xm \in \{0,1\}^{N \times K}$} obtained by stacking a sample of one-hot vectors of the $K$ most recently consumed items. Thus, $\Xm$ is a position-aware interaction matrix that encodes the user's history. The corresponding length-$K$ sequence embedding of the user is $\matr{E}=\Xm^\top\matr{M}$, where $\matr{M}\in\mathbb{R}^{N\times d}$ is an embedding layer of a generic sequential self-attention network.

Positional information is normally injected into this representation via positional embeddings matrix $\matr{P}\in\mathbb{R}^{K \times d}$, producing an updated sequence embedding matrix 
\begin{displaymath}
    \matr{\hat E}=\matr{E+P}.
\end{displaymath}
Define learnable weights 
    \mbox{$\matr{W}_Q\in\mathbb{R}^{d\times d}$},
    \mbox{$\matr{W}_K\in\mathbb{R}^{d\times d}$}, and 
    \mbox{$\matr{W}_V\in\mathbb{R}^{d\times d}$}.
Following the conventional definition of attention weight matrices \mbox{$\matr{Q}=\matr{\hat E}\matr{W}_Q$},
    \mbox{$\matr{K}=\matr{\hat E}\matr{W}_K$}, 
    \mbox{$\matr{V}=\matr{\hat E}\matr{W}_V$}, 
the self-attention layer over the user sequence $\matr{\hat E}$ reads:
\begin{equation}\label{eq:self-attention}
    \operatorname{SA}(\matr{\hat E}) =
    \operatorname{softmax}\left(
      \frac{\matr{\hat E}\matr{W}_Q\matr{W}_K^\top\matr{\hat E}^\top}{\sqrt{d}} \odot\matr{\Omega}
    \right)\matr{\hat E}\matr{W}_V,
\end{equation}
where $\odot$ denotes element-wise product and $\matr{\Omega}$ is the causal masking matrix that prevents future positions from influencing the current prediction; $\operatorname{softmax}$ is applied row-wise.

\subsection{Analytical view of permutation-equivariance}
\label{sec:comparison}
We now focus on dissecting how information flows are organized in self-attention's QKV-blocks internally and analyze the limitations of the conventional approach in terms of modeling the positional awareness. For a cleaner analysis, we begin with a simplified view that omits $\operatorname{softmax}$ and causal masking, as well as drops ad-hoc positional encoding. We then gradually reintroduce these components to investigate their effects and limitations, thereby exposing the underlying inefficiencies and paving the way toward a better solution.

\subsubsection{Attention under sequence permutation}
Consider a single-block sequential self-attention network. Let us expand~\eqref{eq:self-attention} by recalling that a sequence representation is given by \mbox{$\matr{E} = \matr{X}^\top\matr{M}$} (omitting the positional encoding for now), where $\matr{X}$ is an $N\times K$ interaction matrix of some item sequence.
The $K \times d$ output of the (unnormalized) QKV block can then be expressed as:
\begin{equation}\label{eq:YSA}
\left(\matr{Q}\matr{K}^\top\right)\,\matr{V}
=
\left(\matr{X}^\top\matr{M}\,
\matr{W}_Q \matr{W}_K^\top
\matr{M}^\top\matr{X}\right)\,
\matr{X}^\top \matr{M}\,\matr{W}_V.
\end{equation}
This expression makes explicit that the QKV block implements a composition of shared linear operators acting over the interaction matrix $\matr{X}$.

Define a matrix function:
\begin{equation}\label{eq:QKV}
    \matr{Y}(\matr{X}) = \left(\matr{X}^\top\matr{W}\,\matr{X}\right)\matr{X}^\top\matr{M},
\end{equation}
where \mbox{$\matr{W}= \matr{M}\,\matr{W}_Q\matr{W}_K^{\top} \matr{M}^\top\ $} denotes a shared \emph{asymmetric data-independent kernel}. The expression in parenthesis implements similarity calculation between items met in the sequence. Without positional embeddings, it has no positional awareness. Indeed, applying a $K\times K$ permutation matrix $\matr{R}$ over the positional dimension and employing the orthogonality property $\matr{R}\matr{R}^\top=\matr{I}$ gives:
\begin{equation}\label{eq:permutation}
    \matr{Y}(\matr{XR}) = \matr{R}^\top\matr{Y}(\matr{X}).
\end{equation}
Thus, permuting a sequence only leads to a rearrangement of rows in the QKV block output. Obviously, adding the $\operatorname{softmax}$ operator back will have no influence on such rearrangement due to its row-wise nature. This symmetry exposes an undesired property: \emph{sequence permutation does not change the attention weights}.

\subsubsection{Partial position sensitivity with causal masking}

Causal masking ensures that position $k$ can only attend to positions $\{1, \ldots, k\}$, thereby preventing information leakage from the future. While it does not explicitly encode positional dependencies within the available context, it does affect the weighting scheme. Consider two sequences that are permutations of each other, and focus on an intermediate position $k<K$ in each sequence. Due to causal masking, the available contexts will differ: in the first sequence, position $k$ can attend to one subset of past items, while in the second sequence, the same position attends to a different subset. However, within each context, \emph{the attention mechanism weights items based purely on embedding similarity}, treating all past positions symmetrically.

\subsection{Limitations of ad-hoc positional encoding}\label{sec:limitations}

The standard additive positional encoding breaks the permutation symmetry at the embedding level. However, as we show next, it leads to weak, entangled positional modeling.

\paragraph{Gradient dilution across layers.} When positional information is injected only at the input layer, the positional signal must propagate through all subsequent attention blocks via the residual connections. Recent work has shown that this design weakens positional modeling in deep architectures \cite{ke2021diet}. Moving positional information from input embeddings to the attention layer directly improves performance, suggesting that input-level positional signals experience dilution as they propagate through multiple layers \cite{daneshmand2023training,belilovsky2021backward}.

\paragraph{Multi-scale sequential patterns.} Real-world consumption histories may exhibit \emph{multiple overlapping sequential patterns} at different temporal scales. For example, a user may display \emph{short-term trends} by watching a trilogy over $3$ consecutive days along with \emph{weekly periodic patterns} watching action movies every Friday, and all of this while having \emph{long-term preferences} with gradual shift from comedies to dramas over months.
Capturing these diverse patterns can potentially be achieved with attention blocks specialization. Early attention blocks might model short-range dependencies (recent interactions), while deeper blocks capture long-range patterns (periodic behavior or gradual preference shifts). However, uniform positional encoding applied only at the input forces all layers to work with the same positional representation, preventing this hierarchical specialization. Coupled with 
gradient dilution it can potentially weaken the learning of fine-grained positional patterns across network layers.

\paragraph{Position-content interference.} Since each attention block applies non-linear transformations (layer normalization, softmax), the original positional encoding becomes increasingly entangled with item semantics at deeper layers, making it difficult to disentangle position-specific information from content-based representations. Consider a concrete example: a user watches \texttt{Matrix} at position $5$ and \texttt{Inception} at position $50$. With early fusion, the embeddings become
$\tilde{\matr{e}}_5 = \matr{e}_{\text{Matrix}} + \matr{p}_5, \tilde{\matr{e}}_{50} = \matr{e}_{\text{Inception}} + \matr{p}_{50}$
where $\matr{p}_k$ denotes the positional encoding at position $k$. In the first attention block, the model can distinguish that \texttt{Matrix} appeared at position $5$ and \texttt{Inception} at position $50$. However, after several blocks, the attention mechanism mixes these representations:
\begin{equation}
\matr{h}_k^{(L)} = f\big(\sum_{j \leq k} \alpha_{kj} \tilde{\matr{e}}_j\big),
\end{equation}
where $f$ includes normalization and residual connections. At deeper layers, $\matr{h}_k^{(L)}$ contains a weighted combination of multiple positional encodings $\{\matr{p}_j\}_{j \leq k}$, making it ambiguous which position corresponds to which item.

These observations motivate an alternative approach: \emph{explicit position-aware kernels} injected at each attention block, allowing layer-specific positional dependencies that complement rather than replace content-based attention.

\section{Proposed approach: position-aware kernel}

We aim to explicitly model position-dependent weighting through the introduction of a special positional kernel matrix that encodes position-to-position relationships and disentangles them from item semantics.

\subsection{Targeting the position awareness} 
\label{subsection:tpa}

We now introduce a learnable kernel $\matr{C} \in \mathbb{R}^{K \times K}$ that operates solely over positions and multiplicatively modulates attention:
\begin{equation}
    \matr{Y}(\matr{X, C}) = \left(\matr{X}^\top\matr{W}\,\matr{X}\right)\matr{C} \matr{X}^\top\matr{M}.
\end{equation}

The placement of $\matr{C}$ follows from the structure of the bilinear interactions. Treating $\matr{X}$ as the matrix of item row vectors, the term $\matr{X}^\top \matr{W} \matr{X}$ captures item-to-item relations within the semantic space. In contrast, the interaction along the sequence axis corresponds to position-to-position relations. Therefore, inserting $\matr{C}$ between these components introduces a kernel acting purely in the positional dimension, ensuring disentanglement from the semantic item-level interactions governed by $\matr{W}$.

By construction, $\matr{C}$ can implement a recency bias by assigning higher weights to recent positions, or model more complex temporal patterns such as periodicity or decay. The corresponding permutation test~\eqref{eq:permutation} translates into:
\begin{equation}\label{eq:test-pass}
    \matr{Y}(\matr{XR,C}) = \matr{R}^\top\matr{Y}(\matr{X},\,\matr{RCR^\top }) \neq \matr{R}^\top\matr{Y}(\matr{X,C}), 
\end{equation}
which signifies an inherent change in the positional weighting structure beyond mere row reordering.
Equation~\eqref{eq:test-pass} shows that for a fixed learned kernel $\matr{C}$, permutation transformation changes the positional interactions whenever \mbox{$\matr{RCR}^\top \neq \matr{C}$}, so the resulting attention map is no longer permutation-equivariant in the positional dimension. 

The positional kernel $\matr{C}$ provides a multiplicative interaction between content-based attention and structural positional information, enabling the model to distinguish sequences that differ in temporal ordering even when they contain the same set of items. It breaks permutation equivariance structurally inside the attention operator itself, rather than indirectly via additive perturbations of token embeddings. Moreover, by making $\matr{C}$ \emph{learnable and layer-specific}, different attention blocks can model different temporal scales, overcoming the limitations of uniform positional encodings.

\subsection{Learning with positional bilinear operators}

Building on the kernel formulation from the previous section, we now integrate the positional kernel $\matr{C}$ into the standard attention architecture. 
We decompose the kernel matrix as:
\begin{equation}
    \matr{C} = \matr{U}\matr{L},
\end{equation}
where $\matr{U}\in\mathbb{R}^{K\times K}$ and $\matr{L}\in\mathbb{R}^{K\times K}$ are independently parameterized. 
This decomposition ensures $\matr{C}$ retains the capacity of a full interaction matrix, whereas the factors $\matr{U}$ and $\matr{L}$ themselves adopt structural constraints to preserve causality, as detailed in the following subsection. 
Exploiting the structure of the attention operation, we distribute its factors across the softmax boundary to enable independent control over attention scoring and value aggregation. 
The resulting proposed attention mechanism is:
\begin{equation}
    \operatorname{SA}(\matr{E,C}) =
    \operatorname{softmax}\left(
      \frac{\matr{E}\matr{W}_Q\matr{W}_K^\top\matr{E}^\top\matr{U}}{\sqrt{d}} \odot\matr{\Omega}
    \right)\matr{L}\matr{E}\matr{W}_V.
\end{equation}
Specifically, $\matr{U}$ is applied to the attention logits, governing how positional relationships influence attention weights, while $\matr{L}$ is applied to the value vectors, controlling how information is aggregated. 
This formulation explicitly decouples positional modeling from content aggregation. 
The asymmetric incorporation of $\matr{U}$ and $\matr{L}$ allows the model to capture directional temporal effects without imposing symmetry constraints at this stage.

To preserve the sequential structure and prevent information leakage from future positions, we impose structural constraints on the kernel factors. The matrix $\matr{U}$ is constrained to be upper triangular, while the matrix $\matr{L}$ is lower triangular. In the following subsection, we provide an explicit algebraic interpretation of these constraints by analyzing the action of each factor on the sequence representation.

\subsection{Algebraic interpretation of the kernel}

Recall that $\matr{X}$ is a stack of item embedding column vectors ordered chronologically from left (earliest interaction) to right (most recent interaction).

We first analyze the right multiplication by the matrix $\matr{U}$, which appears inside the softmax and therefore directly influences the attention logits. Since $\matr{U}$ is constrained to be upper triangular, the product $\matr{X}^\top \matr{U}$ can be written explicitly as
\begin{equation}
\small
\matr{X}^\top \matr{U}
=
\begin{bmatrix}
\mathbf{x}_1^\top \\
\mathbf{x}_2^\top \\
\vdots \\
\mathbf{x}_K^\top
\end{bmatrix}
\begin{bmatrix}
u_{11} & u_{12} & \cdots & u_{1K} \\
0      & u_{22} & \cdots & u_{2K} \\
\vdots & \ddots & \ddots & \vdots \\
0      & \cdots & 0      & u_{KK}
\end{bmatrix}
=
\begin{bmatrix}
\mathbf{z}_1^\top \\
\mathbf{z}_2^\top \\
\vdots \\
\mathbf{z}_K^\top
\end{bmatrix},
\end{equation}
where each resulting vector is given by
$
\mathbf{z}_k^\top = \sum_{i=1}^{k} u_{ik}\,\mathbf{x}_i^\top.
$
This expression shows that the transformed representation at position $k$ is a weighted combination of item embeddings from the current and all preceding positions only. Consequently, the upper triangular structure of $\mathbf{U}$ guarantees that no information from future interactions contributes to the attention score at any position, ensuring causality at the level of attention logits.

We next consider the left multiplication by the matrix $\mathbf{L}$, which is applied after the softmax and thus affects the aggregation of values. With $\mathbf{L}$ constrained to be lower triangular, we obtain
\begin{equation}
\small
\matr{L}\matr{X}^\top
=
\begin{bmatrix}
\ell_{11} & 0      & \cdots & 0 \\
\ell_{21} & \ell_{22} & \cdots & 0 \\
\vdots & \ddots & \ddots & \vdots \\
\ell_{K1} & \cdots & \ell_{K,K-1} & \ell_{KK}
\end{bmatrix}
\begin{bmatrix}
\mathbf{x}_1^\top \\
\mathbf{x}_2^\top \\
\vdots \\
\mathbf{x}_K^\top
\end{bmatrix}
=
\begin{bmatrix}
\mathbf{y}_1^\top \\
\mathbf{y}_2^\top \\
\vdots \\
\mathbf{y}_K^\top
\end{bmatrix},
\end{equation}
with
$
\mathbf{y}_k^\top = \sum_{i=1}^{k} \ell_{ki}\,\mathbf{x}_i^\top.
$
Here, the representation aggregated at position $k$ depends exclusively on the current and previous positions, which preserves the sequential ordering during value propagation. Together, the asymmetric application of $\matr{U}$ and $\matr{L}$ yields a directional bilinear interaction over positions: $\matr{U}$ controls how past positions influence attention weights, while $\matr{L}$ governs how information from past positions is accumulated in the final representation.

This construction provides a clear algebraic interpretation of the proposed bilinear form and explains how causality is preserved despite allowing fully learnable and asymmetric position-to-position interactions.

\subsection{Implementation}

We conducted several preliminary ablation studies outlined in Section~\ref{subsec:ablation} over different parameterizations of $\matr{C}$, focusing on the structural choices for its factors $\matr{U}$ and $\matr{L}$. Based on these experiments, we adopt the following configuration as the final implementation.

\begin{figure}[t]
    \centering
    \includegraphics[width=1\linewidth]{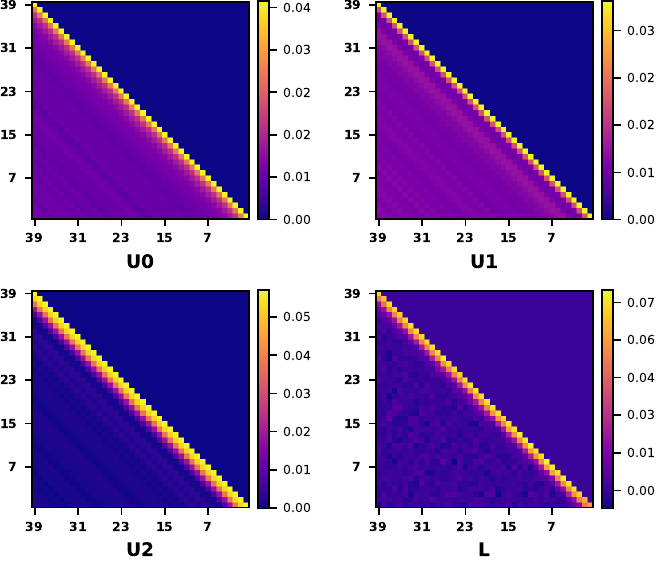}
    \caption{Visualization of trained $\matr{U}$ (per layer) and $\matr{L}$ matrices on the \textit{ml-1m} dataset.}
    \label{fig:u_l}
\end{figure}

The matrix $\matr{U}$ is parameterized as an upper-triangular banded Toeplitz matrix, where all elements along the same diagonal share the same value. Each attention layer is associated with its own matrix $\matr{U}$, resulting in $K$ additional learnable parameters per attention layer. The matrix $\matr{L}$ is parameterized as a fully learnable lower-triangular matrix without additional structural constraints. It is \emph{shared across all attention layers}. The structure of the learned weights is exemplified in Figure~\ref{fig:u_l}.

\paragraph{Complexity analysis.}
For a model with $B$ attention blocks, the position-aware kernel introduces $BK + \frac{K(K+1)}{2}$ learnable parameters.
In typical recommender deployments, $B \ll K \ll d$, so the added overhead is comparable to or even smaller than in the additive approach that requires $Kd$ parameters. Furthermore, since $K$ is typically limited to a few dozen positions in practical applications and only a small number of attention blocks $B$ are used, the parameter overhead remains negligible relative to matrices $\matr{W}_Q, \matr{W}_K, \matr{W}_V$, which together contribute $3Bd^2$ parameters.
Our kernel incurs additional $\mathcal{O}(BK^2d)$ cost of computing attention scores, which increases computation by a factor of $\mathcal{O}(BK)$ relative to elementwise addition requiring only $\mathcal{O}(Kd)$ operations. Leveraging the special structure of our kernel can help to reduce this factor to $\mathcal{O}(B\log K)$.

\begin{figure*}[!htbp]
    \centering
    \includegraphics[width=1\linewidth]{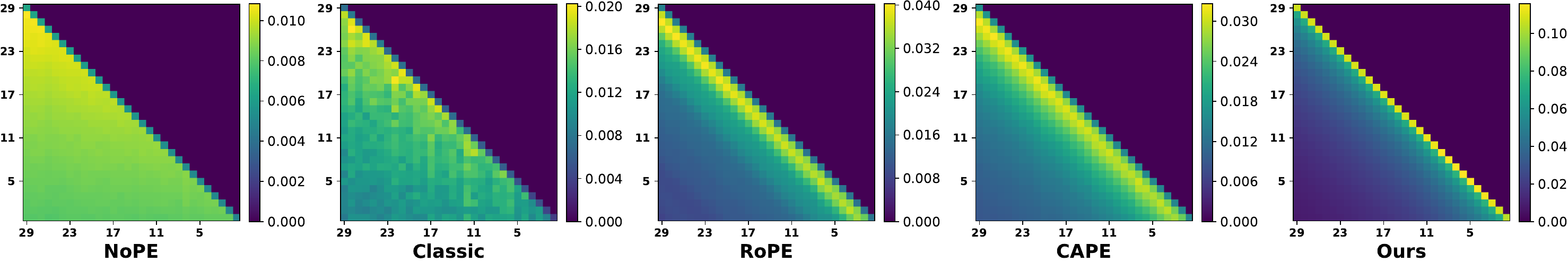}
    \caption{Heatmap of attention matrices of models on the \textit{Y-listens} dataset}
    \label{fig:attenntion}
\end{figure*}

\section{Related works}\label{sec:related}

\paragraph{Positional Encoding in Transformers.}
Early approaches employ additive absolute positional embeddings injected at the input level \cite{vaswani2017attention}. Subsequent work introduced operator-level positional modeling, designed to adjust or modulate the attention mechanism with positional signals. This includes relative position embeddings added to attention logits \cite{shaw2018self}, learned relative bias terms as in T5 \cite{raffel2020t5}, distance-based attention biases such as ALiBi \cite{press2022alibi} that penalize attention to distant positions via fixed linear decay, rotary position embeddings (RoPE) that encode relative offsets through query-key rotations \cite{su2021roformer}, and kernelized positional encodings (KERPLE) \cite{chi2022kerple}. 
While these methods provide effective mechanisms for modeling relative position, they often impose fixed or low-dimensional functional forms on positional effects (e.g., monotonic linear decay in ALiBi or scalar kernels in KERPLE), limiting their ability to capture complex, non-monotonic positional interactions. Additionally, some approaches tightly couple positional information with content representations, as in RoPE, which may cause the position-content entanglement issues identified in Section~\ref{sec:limitations}.

\paragraph{Positional Modeling in Sequential Recommendation.}
User interaction histories are typically short, while the item vocabulary is several orders of magnitude larger than NLP token vocabularies. As a result, many NLP-oriented positional encodings are optimized for length extrapolation and memory efficiency rather than for modeling fine-grained positional effects within short sequences, limiting their practical relevance as baselines in recommender systems.
Early Transformer-based recommender systems, such as \sasrec\cite{kang2018sasrec}, directly adopt additive positional embeddings from NLP. Subsequent works incorporate time-interval modeling (TiSASRec \cite{li2020tisasrec}), explore asymmetrical (PSA \cite{wu2022progressive}) and multi-temporal (MEANTIME \cite{cho2020meantime}) approaches. More recent work incorporates richer positional signals through context-aware position encoding (CAPE \cite{yuan2025cape}). Importantly, \citet{li2024positional} distinguish \emph{positional order} (sequential position within a session) from \emph{temporal footprint} (actual timestamp gaps), showing these require separate modeling in recommendation.

Existing approaches primarily inject positional information at the input level, leaving the self-attention operator structurally unchanged and limiting control over how positional effects influence attention distributions across different network depths.

In contrast, our work introduces structured, learnable position-to-position transformations directly inside the attention computation.
\begin{table}[!b]
    \centering
    \small
    \setlength{\tabcolsep}{3pt}
    \begin{tabular}{l|rrrrr}
\toprule
\textbf{Dataset} & \textbf{\#Users} & \textbf{\#Items} & \textbf{\#Events} & \textbf{Avg. Seq. Len} & \textbf{Density} \\
\midrule
ml-1m    & 6,040   & 3,706    & 1M    & 152.5   & 4.47\% \\
beauty   & 22,363  & 12,101   & 198K  &   8.3   & 0.07\% \\
gowalla  & 27,516  & 173,511  & 2.6M  &  92.9   & 0.06\% \\
yelp     & 42,429  & 137,039  & 2.2M  &  52.3   & 0.04\% \\
zvuk     & 5,000   & 604,543  & 46M   & 9303.5  & 1.54\% \\
Y-likes    & 5,037   & 34,481   & 635K  & 156.1   & 0.37\% \\
Y-listens  & 9,128   & 353,262  & 45.5M & 8335.5  & 1.41\% \\
\bottomrule
\end{tabular}
    \caption{Statistics of the evaluation datasets}
    \label{tab:dataset_statistics}
\end{table}
Unlike existing relative-bias (ALiBi) or kernel-based (KERPLE) approaches that model positional effects as scalar functions of distance, 
our formulation enables flexible interactions between positions while preserving causal structure. This design is well-suited for the short-sequence, large-vocabulary setting of sequential recommendations, where fine-grained, dataset-specific positional effects are critical for accurate next-item prediction.

\section{Experimental setup}\label{sec:experiments}


\subsection{Datasets}\label{subsec:settings}

We evaluate our model on seven datasets from diverse domains. The \textbf{ml-1m} dataset \cite{harper2015movielens} is a widely used benchmark containing one million movie ratings from the MovieLens recommendation service. The \textbf{beauty} dataset 
\cite{ni2019justifying} consists of user interactions with beauty products collected from Amazon’s e-commerce platform. The \textbf{gowalla} dataset \cite{cho2011friendship}, originating from the location-based social network Gowalla, includes user check-ins at various locations. The \textbf{yelp} dataset \cite{asghar2016yelp} contains user ratings and reviews of local businesses from the Yelp platform. The \textbf{zvuk} dataset \cite{shevchenko2024variability} is a large-scale proprietary collection of music streaming interactions from a major Russian music streaming service. In addition, we use two subsets of the Yambda music platform \cite{ploshkin2025yambda}: \textbf{Y-likes}, which contains user likes of music tracks, and \textbf{Y-listens}, which contains user streaming listen events. All datasets undergo 5-core filtering, retaining only users and items with at least five interactions.

We perform a global temporal split with training interactions containing up to the 95th time percentile, validation -- from the 95th to 97th percentile, and test -- the remaining 3 percents of the most recent interactions. For validation and testing, we use \textbf{successive evaluation} \cite{gusak2025time}.
The \textbf{zvuk}, datasets are significantly larger than the others, so we subsample the top 5000 users with the longest interaction sequences to enable feasible experimentation. Dataset statistics are summarized in Table~\ref{tab:dataset_statistics}.

\subsection{Baselines}

We compare our method against four baselines, including recent strong approaches:
    \paragraph{NoPE:} No positional embeddings, serves as a purely content-based attention baseline.

\begin{table}
\centering
\small
\setlength{\tabcolsep}{5pt}
\begin{tabular}{l r | c c c c c}
\toprule
 & \textbf{Metrics} & \textbf{NoPE} & \textbf{Classic} & \textbf{RoPE} & \textbf{CAPE} & \textbf{Ours} \\
\midrule
\multirow{3}{*}{\rotatebox{90}{\scalebox{0.85}{\textbf{ml-1m}}}} 
    & NDCG@10 & 0.0715 & \underline{0.0758} & \underline{0.0817} & 0.0813 & \textbf{0.0828} \\
    & HR@10   & 0.1361 & 0.1452 & \underline{0.1549} & 0.1544 & \textbf{0.1564} \\
    & COV@10  & 0.7882 & 0.7339 & 0.7931 & \underline{0.7939} & \textbf{0.8061} \\
\midrule
\multirow{3}{*}{\rotatebox{90}{\scalebox{0.85}{\textbf{beauty}}}} 
    & NDCG@10 & 0.0316 & 0.0328 & 0.0347 & \underline{0.0377} & \textbf{0.0394} \\
    & HR@10   & 0.0571 & 0.0627 & 0.0650 & \underline{0.0670} & \textbf{0.0683} \\
    & COV@10  & 0.6302 & 0.5079 & \underline{0.6347} & \textbf{0.6376} & 0.6201 \\
\midrule
\multirow{3}{*}{\rotatebox{90}{\scalebox{0.85}{\textbf{gowalla}}}} 
    & NDCG@10 & \underline{0.0477} & 0.0434 & 0.0454 & 0.0433 & \textbf{0.0485} \\
    & HR@10   & \underline{0.0795} & 0.0726 & 0.0764 & 0.0737 & \textbf{0.0803} \\
    & COV@10  & 0.3679 & 0.3319 & \underline{0.3809} & \textbf{0.4137} & 0.3708 \\
\midrule
\multirow{3}{*}{\rotatebox{90}{\scalebox{0.85}{\textbf{yelp}}}} 
    & NDCG@10 & 0.0158 & \underline{0.0164} & \underline{0.0164} & \underline{0.0164} & \textbf{0.0177} \\
    & HR@10   & 0.0332 & \underline{0.0338} & 0.0336 & 0.0328 & \textbf{0.0359} \\
    & COV@10  & 0.1804 & 0.1593 & \underline{0.1845} & \textbf{0.1881} & 0.1672 \\
\midrule
\multirow{3}{*}{\rotatebox{90}{\scalebox{0.85}{\textbf{zvuk}}}} 
    & NDCG@10 & 0.0083 & 0.0081 & \textbf{0.0106} & \underline{0.0100} & 0.0094 \\
    & HR@10   & 0.0113 & 0.0109 & \textbf{0.0141} & \underline{0.0134} & 0.0127 \\
    & COV@10  & 0.0560 & 0.0496 & \underline{0.0618} & 0.0505 & \textbf{0.0634} \\
\midrule
\multirow{3}{*}{\rotatebox{90}{\scalebox{0.85}{\textbf{Y-likes}}}} 
    & NDCG@10 & 0.0147 & 0.0145 & 0.0148 & \underline{0.0149} & \textbf{0.0151} \\
    & HR@10   & 0.0277 & \underline{0.0288} & 0.0285 & \textbf{0.0293} & 0.0286 \\
    & COV@10  & \textbf{0.6695} & 0.5725 & 0.4882 & 0.6120 & \underline{0.6523} \\
\midrule
\multirow{3}{*}{\rotatebox{90}{\scalebox{0.85}{\textbf{Y-listens}}}} 
    & NDCG@10 & 0.0095 & 0.0095 & \underline{0.0114} & 0.0107 & \textbf{0.0123} \\
    & HR@10   & 0.0146 & 0.0152 & \underline{0.0175} & 0.0168 & \textbf{0.0191} \\
    & COV@10  & 0.0952 & 0.0752 & \underline{0.0957} & \textbf{0.0969} & 0.0858 \\
\bottomrule
\end{tabular}
\caption{Performance comparison across all datasets. Bold scores are the best on the dataset for the given metric, underlined scores are the second best.}
\label{tab:results}
\end{table}

    \paragraph{Classic:} Additive positional embeddings, represents the conventional positional encoding scheme in sequential recommendation.
    \paragraph{RoPE:} Rotary positional embeddings, which encode relative positional information through query-key rotations, adapted to the recommendation task \cite{wei2025rotate}.
    \paragraph{CAPE:} Contextual positional encoding, adapts positional representations using local item context \cite{yuan2025cape}.

\paragraph{}
All models are trained on the single-head \sasrec backbone with full Cross-Entropy loss, which has been shown to provide state-of-the-art convergence and performance in sequential recommendation tasks \cite{klenitskiy2023turning} compared to existing Binary Cross-Entropy variants. Moreover, it was also adapted for large item catalogs based on scalable computation algorithms \cite{mezentsev2024scalable,gusak2024rece}. We also adopt the SCE approach from \cite{mezentsev2024scalable} to effectively deal with large datasets. 
Hyperparameter optimization is performed using Optuna with the TPE sampler algorithm, running 60 optimization trials per dataset and model variant.
Each trial trains for up to 300 epochs with early stopping if NDCG@10 on the validation set does not improve for 10 consecutive epochs.

\section{Results}
\label{sec:results}

Table~\ref{tab:results} reports the performance comparison between the proposed position-aware kernel attention and standard positional encoding baselines across seven benchmarks. The results consistently support our hypothesis that modeling positional dependencies directly inside the attention operator yields stronger sequential representations than input-level additive encodings.

Our method achieves the best NDCG@10 on six out of seven datasets and the best HR@10 on five datasets. The improvements are most pronounced on \textit{ml-1m} and \textit{Y-listens}, where long interaction histories and strong recency effects emphasize the importance of explicit temporal modeling. These gains align with the analysis in Section~2.3, indicating that kernel-based modulation of attention mitigates positional signal dilution in deep architectures.

Coverage (COV@10) remains competitive across all datasets, with the highest values achieved on \textit{ml-1m} and \textit{zvuk}. No systematic trade-off between ranking quality and coverage is observed.
The only dataset where our method does not outperform RoPE in ranking metrics is \textit{zvuk}, which contains extremely long music streaming sequences with repetitive and session-level patterns. In such settings, rotary encodings emphasizing relative positional offsets appear particularly effective. Nevertheless, our approach attains the highest COV@10 on this dataset.

Overall, the results demonstrate that disentangling positional structure from item semantics within the attention operator provides a robust and expressive mechanism for sequential modeling. The heterogeneous behavior across datasets further suggests that different positional schemes may capture complementary temporal signals, motivating the combined experiments in the next subsection.

\subsection{Combined approaches}
\label{Combined}

\begin{table}
\centering
\small
\setlength{\tabcolsep}{4pt}
\begin{tabular}{l| c c }
\toprule
\textbf{Metrics} & \textbf{CAPE+Our (like)} & \textbf{RoPE+Our (zvuk)}  \\
\midrule
NDCG@10 &  0.0153 +2.4\%  & 0.0106 +0.1\%\\
HR@10   &  0.0298 +1.9\% & 0.0140 -0.7\%\\
COV@10  &  0.6347 +3.7\%  & 0.0616 -0.4\%\\
\bottomrule
\end{tabular}
\caption{Metrics of combined approaches and improvement compared to model without our approach}
\label{tab:additional_results}
\end{table}

It is interesting to highlight two particular datasets, on which our approach struggles to outperform other baselines: \emph{zvuk} and \emph{likes}. We hypothesize that behavioral patterns in these datasets have a complex nature that cannot be fully captured by the selected kernel structure in our approach that performed well on other datasets. On the other hand, our approach does not interfere with alternative positional encoding schemes as it applies outside the QK kernel of the attention mechanism. Guided by this observation, we integrate our positional kernel to make it work along the other encodings to verify whether it can boost their performance further.   
We attach our positional kernel to the best pretrained models and fine-tune only the bilinear positional form. As shown in Table \ref{tab:additional_results}, our method improves the CAPE model on the Y-likes dataset providing the best-of-all quality comparing to all other baselines. At the same time, no improvement is observed when combined with RoPE on the zvuk dataset. This result shows that, while promising, integrating our positional kernel with other positional encoding schemes requires a deeper analysis of potential interference of the positional patterns learned by different types of the encodings.

\subsection{Attention weights and positional kernel}\label{subsec:visualization}

\begin{table}[t]
\setlength{\tabcolsep}{4pt}

\centering
\begin{tabular}{l | cc | cc | cc}
\toprule
\textbf{model} 
& \multicolumn{2}{c}{\textbf{ml-1m}} 
& \multicolumn{2}{c}{\textbf{beauty}} 
& \multicolumn{2}{c}{\textbf{gowalla}} \\
\textbf{type}
& per-l. & shared 
& per-l. & shared 
& per-l. & shared \\
\midrule
T-T 
& 0.0725 & 0.0828 
& 0.0322 & 0.0359 
& 0.0418 & 0.0430 \\

T-F 
& \textbf{0.0828} & 0.0818 
& \textbf{0.0394} & 0.0378 
& \textbf{0.0485} & 0.0468 \\

F-T 
& 0.0796 & 0.0789 
& 0.0351 & 0.0331 
& 0.0418 & 0.0415 \\
\bottomrule
\end{tabular}
\caption{Ablation study comparing parameterization strategies for bilinear factors $\mathbf{U}$ and $\mathbf{L}$ with and without shared weights (NDCG@10). 
T-T denotes Toeplitz--Toeplitz, T-F denotes Toeplitz--Full, and F-T denotes Full--Toeplitz parameterization. 
Bold scores are the best on the dataset.}
\label{tab:as}
\end{table}

\subsection{Ablation studies}
\label{subsec:ablation}

To analyze the impact of structural constraints and weight sharing in the bilinear positional factors $\matr{U}$ and $\matr{L}$, we conduct an ablation study. All variants are evaluated using NDCG@10, with results reported in Table~\ref{tab:as}. For each configuration, hyperparameters are tuned independently to ensure a fair comparison.

We consider three structural parameterization strategies for the bilinear kernel: Toeplitz--Toeplitz, Toeplitz--Full, and Full--Toeplitz. For each strategy, we further compare layer-wise (per-l.) and shared parameterizations.

The \textbf{Toeplitz--Toeplitz} configuration represents a fully structured kernel with translation-invariant positional interactions and minimal parameter overhead. While the layer-wise variant performs comparably to the classic baseline, weight sharing consistently improves results across all datasets, indicating that a coherent positional decay pattern shared across layers serves as an effective inductive bias under strong structural constraints.

The \textbf{Toeplitz--Full} configuration combines a structured upper-triangular Toeplitz factor with a fully learnable lower-triangular factor. This variant achieves the best performance across all datasets in the layer-wise setting, demonstrating that structured positional bias inside the attention logits, together with flexible value aggregation, provides a favorable balance between regularization and expressiveness. In contrast, sharing the bilinear factors leads to a consistent performance drop, suggesting that layer-wise specialization stabilizes the flexible component.

The \textbf{Full--Toeplitz} configuration follows a similar trend. Although it improves over the fully structured Toeplitz--Toeplitz model, it remains inferior to the Toeplitz--Full variant, and weight sharing again slightly degrades performance.

Overall, the ablation results show that both the placement of structural constraints and the decision to share parameters across layers critically influence performance. The Toeplitz--Full model with layer-wise bilinear factors consistently achieves the highest NDCG@10 across all datasets and is therefore adopted as the default instantiation in subsequent experiments (Table~\ref{tab:results}).

In the next subsection, we provide qualitative visualizations of attention maps and the corresponding bilinear factors $\mathbf{U}$ and $\mathbf{L}$ to further analyze the learned positional patterns.

Figure~\ref{fig:attenntion} demonstrates that the proposed model learns a strictly ordered and interpretable temporal attention structure on the \textit{listens} dataset. The attention weights exhibit a monotonic decay starting from the current position, assigning the highest importance to the most recent interaction and gradually decreasing with temporal distance. This behavior indicates that the positional kernel successfully enforces a clear recency-oriented hierarchy over historical interactions.

This structured pattern directly follows from the explicit disentanglement between content-based item interactions and position-to-position dependencies in our formulation. While semantic similarity is modeled through standard query–key attention, positional influence is governed independently by the bilinear kernel operating purely in the position space. As a result, positional ordering is imposed inside the attention operator itself, leading to a stable and interpretable temporal attention profile.

In contrast, existing positional encoding schemes fail to produce such a clear hierarchy. The NoPE baseline yields an almost uniform lower-triangular structure, indicating aggregation over the entire past without positional discrimination. Classic additive embeddings introduce only a mild directional bias, while the attention distribution remains relatively diffuse.

RoPE and CAPE produce very similar patterns with smoothly decaying attention over past positions; however, in both cases the peak attention is shifted toward recent history rather than the current item, reflecting a focus on relative positional relationships instead of strict recency.

This qualitative distinction aligns with the quantitative results, where the proposed model achieves the largest performance gains on the \textit{listens} dataset (Table~\ref{tab:results}), confirming that the learned positional kernel effectively captures the long-range sequential dynamics of the data.

Figure~\ref{fig:u_l} visualizes the learned bilinear factors $\matr{U}$ and $\matr{L}$ on the \textit{ml-1m} dataset, highlighting the layer-specific specialization of the kernel. The matrices $\matr{U}$ exhibit distinct patterns across depths: the first layer shows a relatively uniform attention distribution, capturing broad global dependencies; the second layer focuses on items at a specific temporal lag (approximately 5 to 7 positions), reflecting mid-range periodicity; and the final layer demonstrates a strong recency bias, prioritizing recent interactions. 
This progression validates the claim in Section~\ref{subsection:tpa} that different attention blocks can model different temporal scales, enabling the architecture to adaptively capture multi-scale sequential dynamics. 
The shared $\matr{L}$ matrix complements this by aggregating historical context, functioning as a history-aware operator across all layers.

\begin{figure}
    \centering
    \includegraphics[width=1\linewidth]{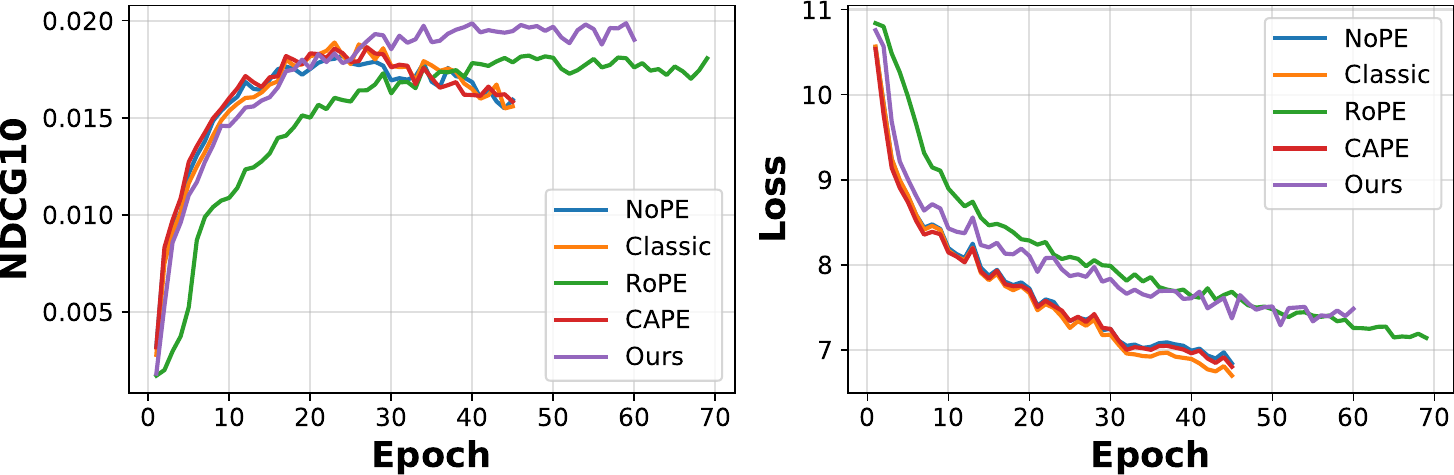}
    \caption{Training dynamics of models on \textit{yelp} dataset }
    \label{fig:train}
\end{figure}

Figure~\ref{fig:train} illustrates the training dynamics on the Yelp dataset in terms of both NDCG@10 and training loss. NoPE, Classic and CAPE methods rapidly converge during the first epochs and then reach a plateau, after which models start to overfit, as reflected by the stagnating or degrading NDCG@10 despite decreasing loss, and training is terminated by early stopping. In contrast, the proposed model and RoPE exhibit more stable training behavior, characterized by a smoother loss decay and consistent ranking performance without a clear degradation trend. Moreover, our approach achieves the highest and most stable NDCG@10 values in the later training stages while maintaining competitive loss dynamics.

Overall, these visualizations demonstrate that the proposed position-aware interaction kernel provides a flexible and interpretable mechanism for modeling positional dependencies, automatically adapting both short-term and long-term regimes according to dataset characteristics.

\section{Conclusion}

In this work, we proposed a position-aware kernelized self-attention mechanism that integrates a learnable kernel over sequence positions directly into the attention operator. This formulation captures explicit position-to-position dependencies without future leakage and disentangles spatial information from item semantics, thereby structurally resolving the permutation-equivariance nature of the standard self-attention. Extensive experiments on both sparse and dense next-item prediction benchmarks show consistent improvements over architectures equipped with competing positional encoding schemes.


\bibliographystyle{named}
\bibliography{ijcai26}

\end{document}